\documentclass[twocolumn]{aastex63}

\usepackage[T1]{fontenc}
\usepackage{aas_macros}
\usepackage{natbib}
\usepackage{times}

\newcommand{\package}[1]{\textsl{#1}}

\newcommand{\msun}{\mbox{M$_\odot$}}

\newcommand{\be}{\begin{equation}}
\newcommand{\ee}{\end{equation}}
\newcommand{\bea}{\begin{eqnarray}}
\newcommand{\eea}{\end{eqnarray}}

\received{2020 October 22}
\revised{2021 March 6}
\accepted{2021 March 12}

\submitjournal{ApJ Letters}
\shorttitle{When the Peas Jump around the Pod}
\shortauthors{Chevance, Kruijssen \& Longmore}

\usepackage{amsmath}

\begin{document}\sloppy\sloppypar\raggedbottom\frenchspacing

\title{\vspace{-12mm}When the Peas Jump around the Pod:\\ How Stellar Clustering Affects the Observed Correlations between Planet Properties in Multi-Planet Systems}

\correspondingauthor{M\'{e}lanie~Chevance}
\email{chevance@uni-heidelberg.de}

\author[0000-0002-5635-5180]{M\'{e}lanie~Chevance}
\affil{Astronomisches Rechen-Institut, Zentrum f\" ur Astronomie der Universit\"at Heidelberg, M\"onchhofstra\ss e 12-14, D-69120 Heidelberg, Germany}

\author[0000-0002-8804-0212]{J.~M.~Diederik~Kruijssen}
\affil{Astronomisches Rechen-Institut, Zentrum f\" ur Astronomie der Universit\"at Heidelberg, M\"onchhofstra\ss e 12-14, D-69120 Heidelberg, Germany}

\author[0000-0001-6353-0170]{Steven~N.~Longmore}
\affil{Astrophysics Research Institute, Liverpool John Moores University, IC2, Liverpool Science Park, 146 Brownlow Hill, Liverpool L3 5RF, UK}


\begin{abstract}\noindent
Recent studies have shown that the radii and masses of adjacent planets within a planetary system are correlated. It is unknown how this `peas-in-a-pod' phenomenon originates, whether it is in place at birth or requires evolution, and whether it (initially) applies only to neighboring planets or to all planets within a system. Here we address these questions by making use of the recent discovery that planetary system architectures strongly depend on ambient stellar clustering. Based on \textit{Gaia}'s second data release, we divide the sample of planetary systems hosting multiple planets into those residing in stellar position-velocity phase space overdensities and the field, representing samples with elevated and low degrees of external perturbation, respectively. We demonstrate that the peas-in-a-pod phenomenon manifests itself in both samples, suggesting that the uniformity of planetary properties within a system is not restricted to direct neighbors and likely already exists at birth. The radius uniformity is significantly elevated in overdensities, suggesting that it can be enhanced by evolutionary effects that either have a similar impact on the entire planetary system or favour the retention of similar planets. The mass uniformity may exhibit a similar, but weaker dependence. Finally, we find ordering in both samples, with the planet radius and mass increasing outwards. Despite its prevalence, the ordering is somewhat weaker in overdensities, suggesting that it may be disrupted by external perturbations arising from stellar clustering. We conclude that a comprehensive understanding of the `peas-in-a-pod` phenomenon requires linking planet formation and evolution to the large-scale stellar and galactic environment.
\end{abstract}

\section{Introduction}
\label{sec:intro}
Over the past decade, the number of exoplanetary systems with multiple planets has increased to the point that statistical studies have become possible. The success of the \textit{Kepler} mission \citep{borucki10,borucki11} has been instrumental in achieving this major step. One of the most surprising discoveries resulting from this growing sample of multi-planet systems has been the correlation between the radii \citep{ciardi2013, weiss2018} and masses \citep{millholland2017, wang2017} of neighboring planets within a system. Not only do planets on adjacent orbits have similar properties, but their orbits are also regularly spaced. Taken together, this coherent behaviour has been named `peas in a pod' \citep{weiss2018}. While it is still debated whether this uniformity could be due to observational biases \citep{zhu2020}, recent statistical arguments imply that the effect is astrophysical in nature \citep{murchikova2020,weiss2020}.

The key question is how the peas-in-a-pod phenomenon originates. Empirically, it may manifest itself in two ways, both of which may coexist and may already be present at the time of formation, or appear only after evolution.
\begin{enumerate}
    \item Planetary properties may be correlated between neighboring planets (i.e.\ on adjacent orbits), even though there may be large-scale gradients throughout a planetary system, such that any random pair of planets within a system exhibits less uniformity than neighboring planets.
    \item Planetary properties may be correlated throughout a planetary system, such that the variance in planet properties between systems is greater than within them.
\end{enumerate}
Because we can only observe planetary systems in their present state, it is challenging to determine which of the above two manifestations applies the most generally, when this uniformity sets in, and if there is an evolution between neighbor-based uniformity and system-wide uniformity.

Planetary systems are known to undergo post-birth evolution in terms of their orbital structure and architecture \citep[e.g.][]{kennedy13}, as well as the planetary composition \citep[e.g.][]{berger20}. If the observed uniformity is in place at birth, then it may be possible to distinguish between neighbor-based and system-wide uniformity if a subset of systems could be identified that experienced significant perturbation or evolution. This might also enable determining if the uniformity is in place at birth, or emerges gradually.

In this \textit{Letter}, we consider the recent discovery that planetary system properties depend on the degree of stellar clustering in their large-scale environment \citep{wklc20}. We take the sample of known planetary systems and use the ambient stellar phase space density obtained with \textit{Gaia} \citep{gaia16,gaia18} to divide the sample into low and high ambient stellar phase space densities \citep{wklc20}, which we refer to as planetary systems in the `field' and in `overdensities', respectively. In a series of companion papers, we have investigated the impact of stellar clustering on the distribution of orbital periods and the incidence of hot Jupiters \citep{wklc20} and on planetary multiplicity \citep[relating to the `Kepler dichotomy', \citealt{lissauer11}]{longmore21}, as well as its role in turning sub-Neptunes into super-Earths \citep[i.e.\ driving them across the `radius valley', \citealt{fulton17}]{kruijssen20d}. These recent findings demonstrate that the \textit{current} degree of stellar clustering in position-velocity space has a major impact on the architectures of planetary systems, plausibly through \textit{past} external photoevaporation or dynamical perturbations \citep[e.g.][]{winter20}. By dividing the sample of planetary systems into field and overdensity samples, we aim to carry out a simple experiment to provide further insight into the physics that cause planets to behave as peas in a pod.

The logic of the experiment carried out in this \textit{Letter} is as follows. We use a variety of statistical metrics to quantify the degree of uniformity in planet radii and masses for our field and overdensity samples. Using the observation that stellar clustering affects planetary system architectures (and at least partially seems to do so after the formation of the planetary system has completed\footnote{See e.g.\ the discussions in \citet{wklc20}, on the generation of hot Jupiters and the nature of the phase space overdensities, and in \citet{kruijssen20d}, on the age dependence of the radius valley.}), we consider these samples to reflect conditions of low and elevated perturbation, respectively. Before proceeding further, we should briefly distinguish between the possible ways in which the environment may affect uniformity. External photoevaporation is likely to affect the entire system in a similar way. As such, it would plausibly increase the degree of uniformity or at least maintain it, both between neighbors and across the planetary system at large. By contrast, dynamical perturbations would disrupt or reorder the planetary system architecture. This could break any initial, neighbor-based uniformity while leaving any system-wide uniformity unaffected, or might increase the degree of system-wide uniformity by allowing the retention of similar planets (e.g.\ by unbinding outer planets). While we acknowledge the possibility that other (possibly indirect) connections between environment and uniformity may exist, we do not speculate on such links in this work.

Following the above line of reasoning, an empirical comparison of the peas-in-a-pod phenomenon between overdensities and the field may have the following possible physical implications. If the peas-in-a-pod phenomenon:
\begin{enumerate}
    \item is stronger for the field, this means that the perturbations induced by stellar clustering weaken the uniformity post-formation, which (A) suggests an influence of dynamical perturbations rather than photoevaporation (which should increase uniformity) and also implies that the uniformity is likely to be neighbor-based, such that (B) the variance between systems is similar to that within a single system, but larger than that between adjacent neighbors;
    \item persists equally between the field and overdensities, this either means (A) that the perturbations induced by stellar clustering affect planetary properties in the same way across a system (e.g.\ through photoevaporation), (B) that the uniformity is likely to be system-wide, such that the variance in planet properties is much greater between systems than within systems and a dynamical rearrangement of the planetary system does not change the degree of uniformity, or (C) that the uniformity is not strongly affected by external processes;
    \item is stronger for overdensities, this means that (A) the perturbations induced by stellar clustering strengthen the uniformity and (B) likely do so in a system-wide sense (such that it is not erased by a dynamical rearrangement of the planetary system), either by selecting planets with similar properties to remain, or by causing them to be subject to similar transformations during their formation or evolution (either by photoevaporation or dynamical perturbations).
\end{enumerate}
These scenarios represent hypotheses that need to be tested by future studies directly investigating how the proposed physical mechanisms affect planetary uniformity. For the purpose of this \textit{Letter}, they serve as a practical framework within which a first interpretation of the observations can be made. The results presented in this work provide evidence in favour of the second and third of the above scenarios. The uniformity is at least as pronounced in overdensities as in the field, suggesting that the peas-in-a-pod phenomenon is system-wide (i.e.\ the variance in planet properties between systems is greater than within them), and is plausibly strengthened by the impact of the large-scale stellar environment.

\section{Observations}
\label{sec:obs}

We use stellar and planet properties from the \citet{exoarchive}. In the following, we describe our sample selection.

\begin{figure*}
\centering
\includegraphics[width=\hsize]{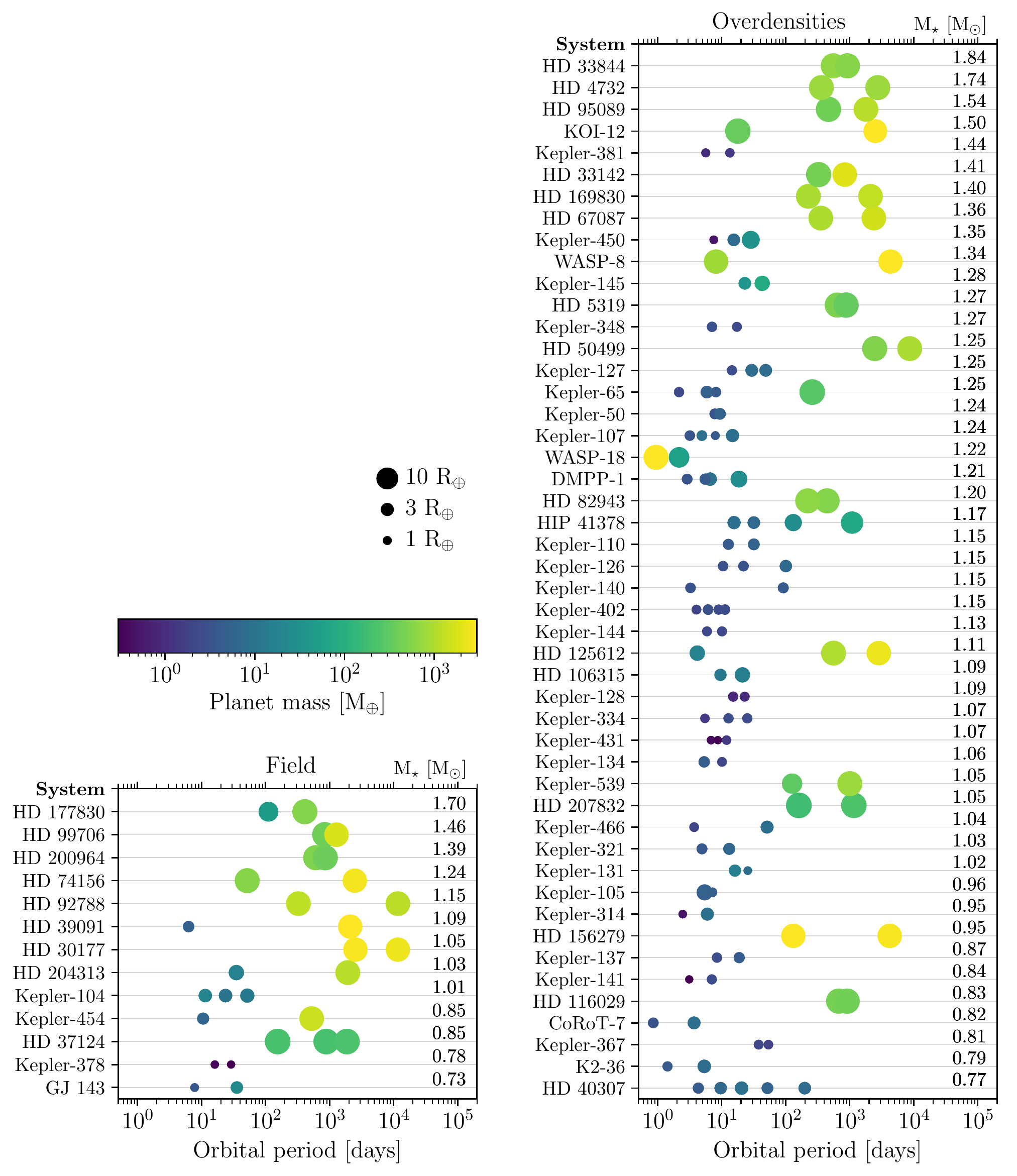}%
\caption{
\label{fig:gallery}
Orbital architectures of the planetary systems in the field (left) and the planetary systems in overdensities (right). The systems are ordered from top to bottom by stellar mass, indicated on the right for each system. The sizes of the dots reflect the planet radii and the colors correspond to the planet masses, as indicated by the color bar. The peas-in-a-pod phenomenon is clearly present, as adjacent planets tend to have similar radii and masses. The visual difference in planet uniformity between the field and overdensity sub-samples at least partially reflects differences in detection method, because the field sample has a larger fraction of planets detected by radial velocity variations. We correct for this mild detection bias through several control experiments. See the text for further discussion.
}
\end{figure*}

\subsection{Stellar clustering}
We investigate whether stellar clustering influences the observed radius and mass uniformity within planetary systems. We build on the analysis presented in \citet{wklc20}, who calculate the relative position-velocity phase space densities of exoplanet host stars. This is done for all host stars of the \citet{exoarchive} that have radial velocities from \text{Gaia}'s second data release \citep{gaia18}, i.e.\ 1522 out of 4141 confirmed exoplanets at the time of sample construction (May 2020). The relative phase space densities are also measured for up to 600 neighboring stars within 40\,pc of each host star. For each host star neighborhood with at least 400 members, the probability $P_{\rm null}$ is calculated that this distribution of phase space densities is drawn from a single Gaussian distribution. In the cases where $P_{\rm null} < 0.05$, a double-lognormal decomposition is performed, which allows us to identify a low-density component and a high-density component, and determine the probability $P_{\rm low}$ (and equivalently $P_{\rm high} \equiv 1-P_{\rm low}$) for the exoplanet host star to be associated with the low-density (and equivalently the high-density) component. In the following, we select planetary systems in the field (defined as having $P_{\rm low} > 0.84$) and in stellar overdensities (defined as having $P_{\rm high} > 0.84$). These represent a total sample of 1033 planets. Other exoplanet host stars are considered to have an ambiguous classification and are therefore not selected here.

\subsection{Planetary system sample}
\label{sec:sample}
In order to both limit the impact of the heterogeneity of the \citet{exoarchive} sample and suppress any trends with host stellar properties, we apply the following additional selection criteria to our sample. We exclude systems younger than 1\,Gyr, that might not yet have stabilized \citep{kennedy13}. Phase space overdensities are likely to disperse dynamically within $\sim4.5$\,Gyr \citep{seabroke2007}. We therefore also omit systems older than 4.5\,Gyr. Additionally, we restrict the host stellar mass range to $M = 0.7 {-} 2.0$\,\msun\ to ensure that the stellar mass distribution is similar in both the low- and high-density sub-samples. Finally, we then keep all multi-planet systems hosting two or more detected exoplanets. This results in a sample of 48 systems in overdensities (116 planets) and 13 systems in the field (28 planets).

Despite these additional selection criteria, it remains possible that there are biases due to differences in detection methods, most prominently between transit and radial velocity surveys. Both techniques have different sensitivity biases, leading to concrete differences in detected planet properties. Additionally, transit surveys provide planet radii, whereas radial velocity surveys provide planet masses. The missing quantity must therefore be estimated using a planet mass-radius relation \citep[we adopt the relation of][]{chen17}, which introduces scatter and may therefore affect the degree of uniformity that we measure. These biases could affect our results, but \textit{only} if the occurrence of the different detection methods differs considerably between the field and overdensity samples. In \S\ref{sec:MC} and \S\ref{sec:bias}, we carry out a set of Monte-Carlo experiments and additional sample divisions through which we quantify the above sources of bias, and find that correcting for detection method bias would strengthen our findings, rather than weaken them.

\autoref{fig:gallery} shows the architectures of all (multiple) planetary systems in our sample, split between systems in the field and systems in overdensities. Visually, it appears that the overdensity systems on average have smaller orbital periods and lower mass planets than field systems. This matches the difference in orbital periods observed when also taking into account single-planet systems in \cite{wklc20}.

While qualitative differences are present between the left and right columns in \autoref{fig:gallery} in terms of the planet uniformity within systems (neighboring planets seem to have more uniform properties in the overdensity sub-sample), we caution against a direct physical interpretation of these visual trends. The field sample has a larger fraction of planets detected by radial velocity measurements than the overdensity sample, which leads to a higher fraction of massive, far-out planets. In order to draw quantitative conclusions regarding the uniformity within each sub-sample, and additionally make a comparison between both, it is necessary to construct control samples through carefully chosen Monte-Carlo experiments.

\subsection{Synthetic control samples}
\label{sec:MC}
Contrary to other studies investigating the peas-in-a-pod phenomenon \citep[e.g.][]{millholland2017,weiss2018}, we do not restrict our sample to a homogeneous planet sample in terms of the detection method. Limiting our sample to the planets only detected by \textit{Kepler} would limit the sample too strongly, preventing us from having statistically significant numbers of systems in the field and overdensity sub-samples. A future, homogeneous sample even larger than the current \textit{Kepler} sample would be necessary to assess the impact of stellar clustering on planet uniformity in a clean experiment, following \citet{millholland2017} and \citet{weiss2018}. Therefore, we instead determine whether the heterogeneous detection methods for different systems can introduce biases in our conclusions. We carry out a number of tests to rule out the importance of several potential sources of bias, and quantify the others.

\begin{deluxetable*}{l @{\hspace{0.5cm}} l l l l l l l}
\tablehead{
Planet sub-sample & Stellar mass & Stellar metallicity & System age & Distance & Number & Number of systems  & Number of systems with \\
& [$\msun$] & [dex] & [Gyr] & [pc] & of systems & with transit detections & radial velocity detections
}
\decimals
\setlength{\tabcolsep}{3pt}
\startdata
% ALL
Field & $1.04_{-0.19}^{+0.35}$ & $0.10_{-0.30}^{+0.18}$ & $3.4_{-0.9}^{+0.5}$ & $63_{-31}^{+144}$ & $13$ & $5$ & $12$ \\
Overdensities & $1.15_{-0.20}^{+0.20}$ & $0.07_{-0.14}^{+0.14}$ & $3.0_{-1.2}^{+1.3}$ & $245_{-186}^{+237}$ & $48$ & $32$ & $27$ \\\hline
Mean uncertainty & 0.08 & 0.07 & 1.1 & 2.7 & & & \\
% RESTRICTED TO CKS CROSS-MATCH
% Field & $1.01_{-0.16}^{+0.00}$ & $-0.20_{-0.00}^{+0.47}$ & $3.7_{-0.0}^{+0.5}$ & $405.0_{-173.0}^{+0.0}$ & $2$ & $2$ & $2$ \\
% Overdensities & $1.15_{-0.13}^{+0.10}$ & $0.04_{-0.07}^{+0.13}$ & $3.3_{-0.9}^{+1.0}$ & $389.0_{-120.0}^{+187.0}$ & $26$ & $25$ & $7$ \\
% RESTRICTED TO CKS CROSS-MATCH, USING CKS STELLAR PARAMETERS
% Field & $0.82_{-0.00}^{+0.24}$ & $-0.42_{-0.00}^{+0.7}$ & $13.2_{-8.8}^{+0.0}$ & $405.0_{-173.0}^{+0.0}$ & $2$ & $2$ & $2$ \\
% Overdensities & $1.08_{-0.09}^{+0.17}$ & $-0.01_{-0.09}^{+0.25}$ & $3.2_{-0.9}^{+1.8}$ & $389.0_{-120.0}^{+187.0}$ & $26$ & $25$ & $7$ \\
\enddata
\caption{Median host stellar properties for each of the two planet sub-samples listed in the first column. Uncertainties indicate the 16th and 84th percentiles of the distributions. The total number of systems in each sub-sample is indicated, as well as the number of systems for which at least one planet has been detected by transit or radial velocity surveys. The final row lists the mean formal uncertainty on each quantity. 
\vspace{-8mm}
}
\label{tab:bias}
\end{deluxetable*}
First, we verify in \autoref{tab:bias} that our sub-samples do not exhibit strong differences in properties of the host star (specifically its mass, metallicity, age, and distance). As the table shows, the median and dispersion of the host star properties are indeed similar in both samples. This conclusion is unaffected when drawing the stellar properties from a different database \citep[e.g.][]{fulton18}. Additionally, the mean formal uncertainties on each of the quantities are smaller than (or equal to) the typical dispersions across each sample, which in turn are much smaller than the ranges spanned by the adopted sample cuts. Taken together, this shows that the similarity between both samples does not result from dilution \citep[also see][]{wklc20}. However, the median distance of the systems differs between the sub-samples, even if the dispersion is so large that both ranges comfortably overlap. In \citet{wklc20}, distance was ruled out as a systematic source of bias between the field and overdensity samples. Finally, we see different proportions of each sub-sample have been characterized by transit or radial velocity surveys. This is a potential source of bias that needs to be addressed.

To enable assessing the statistical significance of our results and additionally be able to control for observational biases, we generate a series of control samples, using two sets of Monte-Carlo realisations. In the first Monte-Carlo experiment, we reshuffle all planets within the \{low, high\}-density sub-samples separately, while keeping the orbits and multiplicity of each system intact. Systems in the field are re-populated with planets randomly drawn from the field sub-sample, whereas systems in overdensities are re-populated with planets randomly drawn from the overdensity sub-sample. In the second Monte-Carlo experiment, we similarly reshuffle all planets while keeping the orbits and multiplicity of each system intact, but now draw the planets from the complete sample, irrespectively of the ambient stellar phase space density. In both Monte-Carlo experiments, we generate $10^4$ realisations of the planetary systems.\footnote{Note that the measured radii and masses are kept constant in these reshuffling experiments. We repeated the experiments when also performing an additional random draw from each measurement and its uncertainty (and propagating it through the \citealt{chen17} mass-radius relation if one of these two quantities has not been measured directly). The resulting Monte-Carlo experiments are statistically indistinguishable from the ones presented in this work.}

The first experiment retains any differences in observational biases between both sub-samples. If any of our findings result from such a bias, it should also appear in this Monte-Carlo set. The second experiment erases any potential observational biases by shuffling planets between both sub-samples. We demonstrate in \S\ref{sec:statistics} that the statistical metrics inferred for both experiments are identical, which gives confidence that our findings are not affected by systematic differences in observational biases between both sub-samples.

Our Monte-Carlo experiments do not provide a test for potential observational biases on the multiplicity and orbital structure within systems, because these are conserved. Radial velocity and transit surveys are typically sensitive to planets on different orbital configurations. This could translate into systematic differences between our two sub-samples, because they contain different fractions of planetary systems detected with either method. While we are not specifically looking at the distribution of orbits or multiplicity, we acknowledge that there might exist a correlation between these observables and the quantities that we do consider here (e.g.\ between orbit and radius, or between orbit and mass). Such a correlation could be caused by differences in formation processes between close-in minor planets and far-out giant planets, and might in turn lead to biases in our conclusions. To address this potential concern, we have repeated our entire analysis for systems in the overdensity sample containing only transit detections (19 systems) and systems in the overdensity sample containing only radial velocity detections (14 systems).\footnote{A similar experiment cannot be performed for the field sample due to the low number of field systems.} This experiment shows that the statistical trends observed in \S\ref{sec:results} are opposite to any trends arising from a detection method bias, and would therefore be strengthened if we would have access to a sufficiently rich, homogeneous sample. As the sample of well-characterized exoplanetary systems continues to grow and future work will be able to repeat our analysis within a sub-sample of systems characterized by a single detection method, these sources of potential concern will be alleviated altogether.

\section{Uniformity of planet properties in field and overdensity systems}
\label{sec:results}

\subsection{Peas in a pod}
\label{sec:peas}
\begin{figure*}
\centering
\includegraphics[width=\hsize]{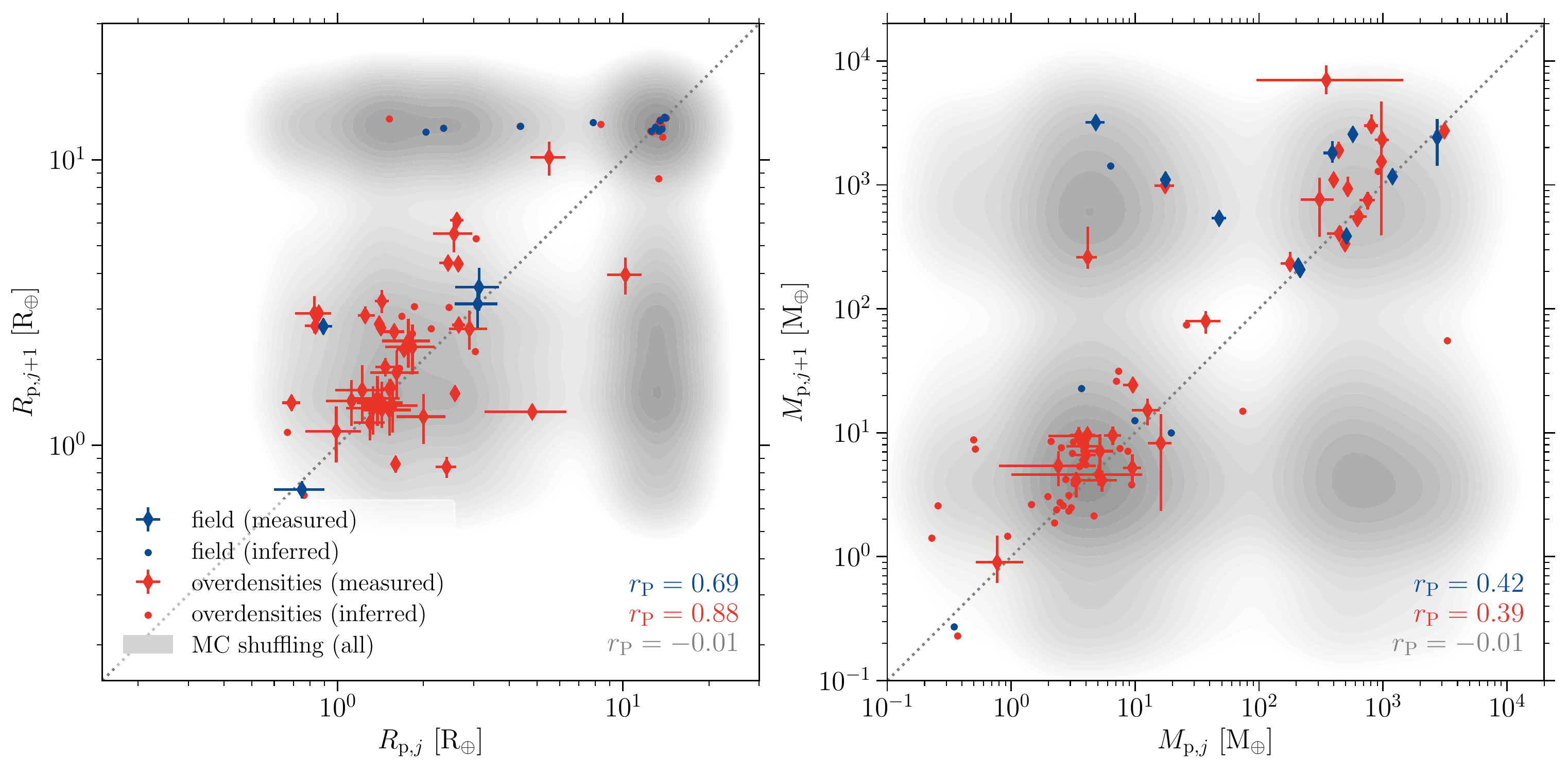}%
\vspace{-2mm}\caption{
\label{fig:peas}
Left: Radius of a planet $R_{{\rm p},j}$ versus the radius of the next adjacent planet $R_{{\rm p},j+1}$. Right: Mass of a planet $M_{{\rm p},j}$ versus the mass of the next adjacent planet $M_{{\rm p},j+1}$. Planets in field systems are shown in blue, whereas planets in overdensity systems are shown in red. Diamond symbols represent pairs of adjacent planets for which the quantity on both axes has been measured directly, with error bars representing the $1\sigma$ uncertainties. Dots represent pairs of adjacent planets for which this quantity has been inferred for at least one of them using the planet mass-radius relation of \citet{chen17}. The grey-shaded density contours represent the results of our Monte-Carlo experiment in which the planets are randomly reshuffled between the combined sample of field and overdensity planets (see the text). In both panels, the Pearson correlation coefficients for each sub-sample (field, overdensity and Monte-Carlo simulation) are indicated in the bottom right corner. Similarities between the properties (especially radius, and also mass) of adjacent planets are observed for both sub-samples. Uniformity in radius is stronger for overdensities (i.e.\ perturbed systems, with $r_{\rm P}=0.88$) than in the field (i.e.\ unperturbed systems, with $r_{\rm P}=0.69$).
}
\end{figure*}

We now discuss whether the observed similarity in properties (mass and radius) of exoplanets within a system \citep{ciardi2013,millholland2017,weiss2018} holds for both sub-samples of planets orbiting field and overdensity stars. \autoref{fig:peas} shows the relation between the sizes and masses of adjacent planets, split between field and overdensity systems. In both cases, we observe a significant correlation between the radii ($R_{j}$, $R_{j+1}$) and masses ($M_{j}$, $M_{j+1}$) of adjacent planets, with Pearson correlation coefficients of $r_{\rm P}=\{0.69; 0.42\}$ for planets in field systems and $r_{\rm P}=\{0.88 ; 0.39\}$ for planets in overdensity systems. This means that we observe the peas-in-a-pod pattern that was found in the \textit{Kepler} data independently of the ambient stellar phase space density \citep{millholland2017, weiss2018} in both of our samples, with a stronger correlation for the radii of adjacent planets than for their masses.\footnote{We caution that the outliers in \autoref{fig:peas} generally represent systems with an inner rocky planet and an outer gas giant with a large period ratio (see \autoref{fig:gallery}). Therefore, it might be possible that these systems have undetected planets in between, which would affect the uniformity signal.} The radius uniformity is stronger for systems in overdensities than for systems in the field. We do not detect qualitative difference in the degree of mass uniformity between both sub-samples.

Our observation suggests that there exists some way of increasing the radius uniformity within systems by stellar clustering. This might arise from several mechanisms. For instance, these systems may be disrupted in such a way that similar planets remain and others are removed. Alternatively, the architectures of the systems may be modified in such a way that the planets become more similar (e.g.\ by moving them closer to the star, evaporating their atmospheres, and driving them down the radius valley, see e.g.\ \citealt{owen13,fulton17,kruijssen20d}). The fact that both the mass and the radius exhibit uniformity suggests that both of these mechanisms may be at play. The fact that the radius uniformity is stronger than the mass uniformity in overdensity systems may suggest that photoevaporation could play an important role \citep[also see][]{kruijssen20d}.

Due to the heterogeneity of detection methods across our sample, some planets only have direct radius (respectively mass) measurements, such that the mass (repectively radius) is inferred from the planet mass-radius relation \citep{chen17}. If we restrict our sample to the planet pairs for which observational measurements are available (see the diamond symbols in \autoref{fig:peas}), the peas-in-a-pod behaviour of planets in overdensity systems may be slightly stronger than for field stars. However, the low-number statistics of the field sub-sample prohibit definitive conclusions.

We compare the observations with the second Monte-Carlo experiment described in \S\ref{sec:MC}, in which planets are shuffled across both sub-samples (field and overdensities). No peas-in-a-pod pattern is observed, and the Pearson correlation coefficients are $r_{\rm P}=-0.01$ for both the radii and the masses of adjacent planets. This quantitatively demonstrates the excess uniformity observed in both sub-samples.

\subsection{Quantitative metrics}
\label{sec:statistics}

We now calculate various statistical metrics to quantify the degrees of uniformity and ordering within planetary systems in the field and in overdensities. In addition, we compare these between our measurements and the Monte-Carlo samples described in \S\ref{sec:MC} and investigate whether our findings might be affected by observational biases.

\begin{figure*}
\centering
\includegraphics[width=0.94\hsize]{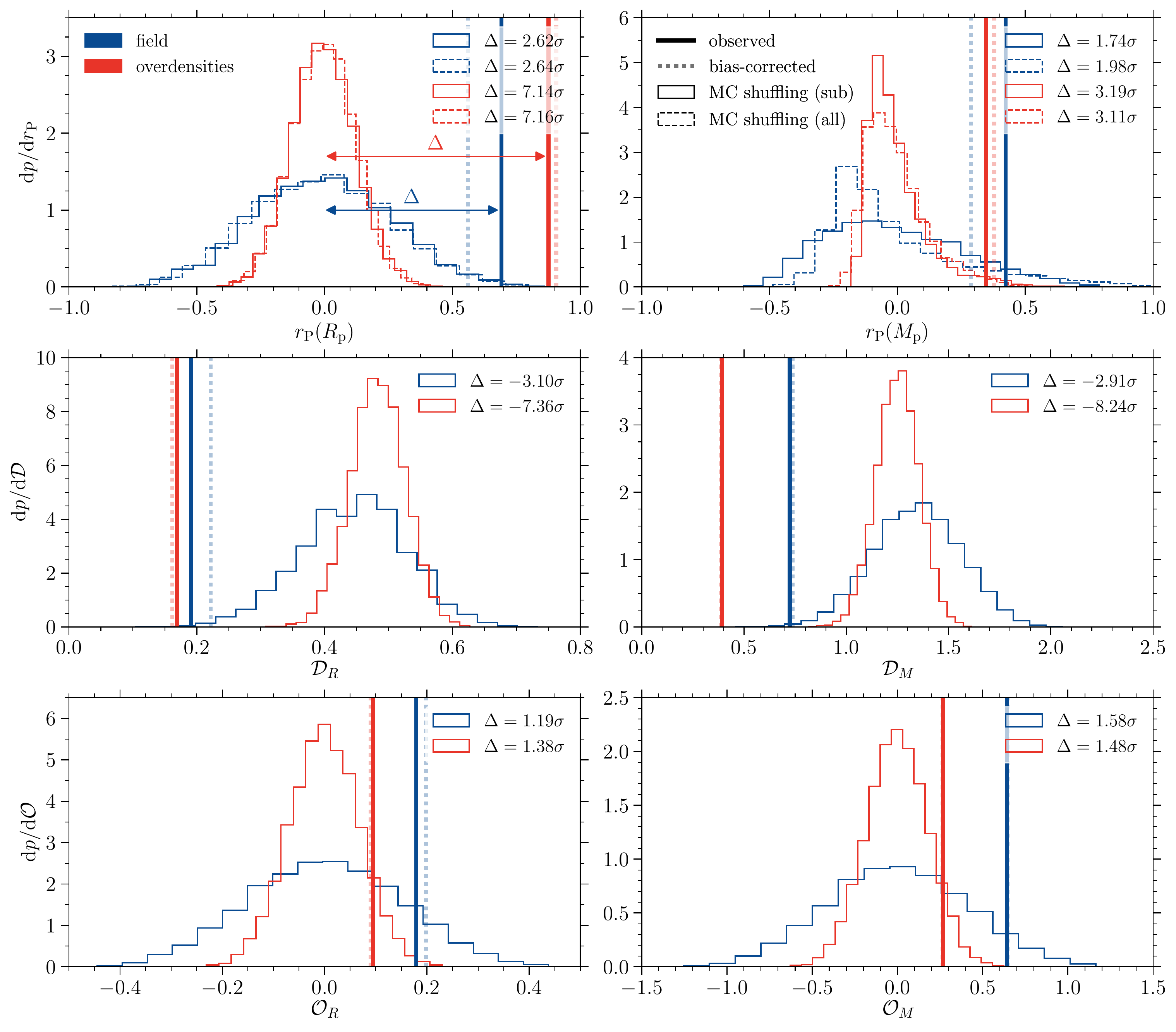}%
\vspace{-2mm}\caption{
\label{fig:metrics}
Quantitative metrics describing the degree of uniformity and ordering for the radii (left) and masses (right) of adjacent planets in multiple systems residing in the field (blue) and in overdensities (red). Each panel shows the observed values as thick vertical lines (with thick dotted lines showing the values corrected for observational method bias, see the text), and the probability distribution functions obtained from the Monte-Carlo control samples (see \S\ref{sec:MC}) as histograms. The solid line histograms show the results of the first Monte-Carlo experiment, in which planets are shuffled within their given sub-sample (field or overdensity). The dashed line histograms (only shown in the top row) show the results of the second Monte-Carlo experiment, in which planets are shuffled across the combined sample. The top row shows the Pearson correlation coefficients (compare \autoref{fig:peas}), with uniformity increasing towards the right. The middle row shows the distance metric defined in \autoref{eq:distance}, with uniformity increasing towards the left. The bottom row shows the ordering metric defined in \autoref{eq:ordering}, with quantities increasing with orbital period towards the right. The difference between the observed values and the medians of the Monte-Carlo samples is indicated in the top right corner of each panel, in units of the standard deviation of the probability distribution function. We find highly statistically significant peas-in-a-pod patterns and moderately significant ordering of both the radii and masses, with stronger uniformity in overdensities.
}
\end{figure*}

In the top row of \autoref{fig:metrics}, we show the observed Pearson correlation coefficients (indicated in \autoref{fig:peas}), as well as the probability distribution functions of both Monte-Carlo experiments. As discussed in \S\ref{sec:peas}, this shows that the uniformity of radii between adjacent planets is stronger than the uniformity in mass, and this uniformity is more pronounced for planetary systems in overdensities. This means that stellar clustering transforms planetary systems in a way that further homogenizes their radii.

The correlation between properties of adjacent planets disappears after reshuffling the planets, irrespectively of whether this is done within or across the sub-samples, as the medians of the probability density functions are close to zero. In addition, we find great similarity between the histograms for both Monte-Carlo experiments (solid and dashed lines) as well as the number of standard deviations their medians differ from the observations ($\Delta$). Because the first Monte-Carlo experiment draws from within each sub-sample (therefore maintaining any detection method and distance differences) and the second Monte-Carlo experiment draws from the entire sample (therefore erasing any detection method and distance differences), their similarity demonstrates that differences in detection method and distance between the field and overdensities do not affect our conclusions. For clarity, the middle and bottom rows of \autoref{fig:metrics} therefore only include the probability distribution functions obtained with the first Monte-Carlo experiment (shuffling within sub-samples), but both experiments yield consistent results. Due to the low-number of field systems, the probability distribution functions for the Monte-Carlo field sample are wider in all panels of \autoref{fig:metrics}. This obstructs a quantitative comparison between the absolute deviation values ($\Delta$) between the field and overdensity sub-samples -- these can only be used to assess the statistical significance of the uniformity. Nonetheless, the absolute value of the Pearson correlation coefficients in the top left panel clearly demonstrates that the uniformity in radius is increased in overdensity systems.

The uniformity in radius and mass between adjacent planets within systems can be described in terms of clustering in the $R_{{\rm p},j}$--$R_{{\rm p},j+1}$ and $M_{{\rm p},j}$--$M_{{\rm p},j+1}$ planes. Following \cite{millholland2017}, we measure the total distance $\mathcal{D}_{R}$ between all adjacent pairs ($N_{\rm pair}$) in the $\log R_{{\rm p},j}$--$\log R_{{\rm p},j+1}$ plane, defined as:
\begin{equation}
    \mathcal{D}_{R} = \sum_{i=1}^{N_{\rm pair}}  \left | \log \frac{R_{{\rm p},j+1,i}}{R_{{\rm p},j,i}} \right | ,
\label{eq:distance}
\end{equation}
where $R_{{\rm p},j,i}$ and $R_{{\rm p},j+1,i}$ are the radii of the inner and outer planet in the pair $i$. The total distance $\mathcal{D}_{M}$ between all adjacent pairs in the $\log M_{j}$-$\log M_{j+1}$ plane is defined in an analogous way. These two quantities are shown in the middle row of \autoref{fig:metrics}. For both quantities and both sub-samples, we measure a significantly stronger clustering (smaller $\mathcal{D}_{R}$ and $\mathcal{D}_{M}$) than for randomized samples from the Monte-Carlo experiment. Based on this metric, the masses also exhibit increased uniformity in overdensities, similarly to the observation for radii in the top left panel.

Recent studies have found that planets within a system tend to be ordered, with larger planets having longer orbital periods \citep{millholland2017, kipping2018}. Following the former of these papers, we calculate the ordering metric $\mathcal{O}_{R}$ for planet radii, defined as:
\begin{equation}
    \mathcal{O}_{R} = \sum_{i=1}^{N_{\rm pair}}  \log \frac{R_{{\rm p},j+1,i}}{R_{{\rm p},j,i}} ,
\label{eq:ordering}
\end{equation}
with an analogous definition for the ordering metric for masses, $\mathcal{O}_{M}$. These two quantities are shown in the bottom row of \autoref{fig:metrics}. In both panels, the ordering of the observed systems is positive, at a moderately statistically significant degree. This means that the radii and masses of planets within a system tend to increase outwards. The ordering may be somewhat stronger for field systems, but this is hard to establish unambiguously given the small number statistics and heterogeneous sample.

\subsection{Influence of detection method bias}
\label{sec:bias}
Finally, we need to consider the possibility that the heterogeneity in detection methods might affect our findings. Systems in the field and in overdensities have been observed in different proportion by different observational techniques: 36\% of the systems in our field sample have at least one planet detected by transit surveys, compared to 66\% in our overdensity sample; 92\% of the systems in our field sample have at least one planet detected by radial velocity surveys, compared to 56\% in our overdensity sample. We use the control experiment in which we divide the overdensity sample in transit-only and radial velocity-only systems to illustrate how a simple, ad-hoc correction for detection method biases would affect our findings.

After dividing the overdensity sample into systems observed only through transit measurements (19 systems) and radial velocity measurements (14 systems), we measure each of the quantities shown in \autoref{fig:metrics} for each detection method sub-sample, providing us with measurements $x_{\rm tr}$ (transits only) and $x_{\rm RV}$ (radial velocities only). Assuming that a sample containing an equal number of transit and radial velocity detections would yield $x=(x_{\rm tr}+x_{\rm rv})/2$, we define the bias of each individual detection method as
\be
\label{eq:bias}
\delta_{\rm tr} = \frac{1}{2}\left(x_{\rm tr}-x_{\rm rv}\right) ~~~{\rm and}~~~
\delta_{\rm rv} = \frac{1}{2}\left(x_{\rm rv}-x_{\rm tr}\right) .
\ee
We weigh these biases by the numbers of pairs that are detected only by transits ($N_{\rm tr}$) or radial velocities ($N_{\rm rv}$) in each of the field and overdensity sub-samples to correct the measurement as
\be
\label{eq:corr}
x\rightarrow x-\left(\frac{N_{\rm tr}}{N_{\rm tr}+N_{\rm rv}}\delta_{\rm tr}+\frac{N_{\rm rv}}{N_{\rm tr}+N_{\rm rv}}\delta_{\rm rv}\right) .
\ee
The results are shown as vertical dotted lines in \autoref{fig:metrics}. Note that this correction automatically accounts for any impact of the high uncertainties on the radii and masses in the radial velocity and transit samples, respectively (see \S\ref{sec:sample}), because we are directly comparing both samples.

Despite the relatively low numbers of systems used in this control experiment, it shows that the differences seen between field and overdensity systems cannot be explained by a detection method bias. Specifically, we find that the radial velocity-only sample exhibits stronger uniformity than the transit-only sample, whereas in our main analysis the strongest uniformity is found in the overdensities (which are dominated by transiting systems). This opposite behaviour means that a correction for detection method bias strengthens our results. Future repeats of our analysis on larger and more homogeneous samples should therefore find even stronger differences between field and overdensity systems.

\section{Discussion}
\label{sec:disc}

We have investigated the origin of the `peas-in-a-pod' phenomenon, i.e.\ the observed uniformity of adjacent planet properties (radii and masses) within planetary systems. Specifically, we have aimed to address whether this uniformity depends on the current degree of stellar clustering in the large-scale stellar and galactic environment of a planetary system. Because enhanced stellar clustering can be a source of elevated perturbations, this is a fruitful division to make when attempting to understand the nature of the observed uniformity. Specifically, quantifying the influence of stellar environment may shed light on whether this uniformity originates during planet formation or is a result of evolution, and whether it is primarily a system-wide property or applies primarily between neighboring planets. To address these questions, we have divided the observed population of multi-planet systems into `field' and `overdensity' sub-samples, containing systems residing in low and high relative phase space densities \citep{wklc20}. We consider these samples to reflect conditions of low and elevated environmental perturbations, respectively. Further sample cuts are made to ensure uniformity of stellar host properties between sub-samples (see \S\ref{sec:obs} and \autoref{tab:bias}). This results in a sample of 13 field systems and 48 overdensity systems.

We show that for radii and masses of adjacent planets, the peas-in-a-pod behaviour persists equally between systems in the field and in overdensities. The fact that systems that are perturbed by stellar clustering still exhibit this uniformity suggests that the correlation of properties between planets is system-wide, rather than between neighbors, and is likely to already exist at formation. Therefore, we conclude that the variance in planet properties between systems is greater than within systems, similar to the conclusion arrived at by \citet{millholland2017} and confirming the predictions of numerical simulations \citep{macdonald2020}.\footnote{We emphasize that this is a statistical statement and should not be interpreted as system-wide uniformity in a strict sense -- even though compact systems of rocky planets often host additional distant giant planets \citep[e.g.][]{bryan19}, the planet properties within the system may be more similar than those across the planet population.} This prevalence of the peas-in-a-pod behaviour is likely explained by the fact that radii and masses are set by the system, so that external perturbations cannot significantly change the uniformity of the system by reordering it. This lends further support to the conclusion drawn by \citet{murchikova2020} that `planets know about the system they formed in'.

Even though the peas-in-a-pod pattern manifests itself both in planetary systems in the field and in overdensities, there are differences between these sub-samples. For planet radii, the uniformity is stronger for systems in overdensities than for systems in the field. This would suggest that external perturbations affect systems globally, either because the potential reordering processes of perturbed systems allows planets with similar radii to remain in the system, or because planets are transformed in a similar way during their formation or evolution (e.g.\ by photoevaporation, dynamical perturbations, or a combination of both). By combining a set of statistical metrics we find tentative evidence that the mass uniformity is also elevated in overdensities, which suggests that multiple perturbation mechanisms may be at play. Finally, both the radii and masses of adjacent planets tend to increase outwards, although slightly less so in the overdensities compared to the field. This suggests that the ordering within systems might be disrupted by external processes. Based on our control experiments, we find that our results are robust against systematic observational (e.g.\ detection method) biases between the field and overdensity sub-samples.

At present, the number of planetary systems for which our analysis can be performed is limited. Nonetheless, we find that the uniformity of planet properties exhibits a statistically significant difference between systems in phase space overdensities and in the field. We expect future studies repeating our work for larger and more homogeneous samples to further refine the trends identified here. Most importantly, our findings demonstrate that a multi-scale approach is necessary to fully understand how planetary systems form and evolve, which takes into account the large-scale stellar environment.

\vspace{-2mm}
\acknowledgments
The authors thank the anonymous referee for a timely and helpful report, and Guiseppina Micela for helpful comments on the manuscript. M.C.\ and J.M.D.K.\ gratefully acknowledge funding from the Deutsche Forschungsgemeinschaft (DFG, German Research Foundation) through an Emmy Noether Research Group (grant number KR4801/1-1) and the DFG Sachbeihilfe (grant number KR4801/2-1). J.M.D.K.\ gratefully acknowledges funding from the European Research Council (ERC) under the European Union's Horizon 2020 research and innovation programme via the ERC Starting Grant MUSTANG (grant agreement number 714907). This research made use of data from the European Space Agency mission \textit{Gaia} (\href{http://www.cosmos.esa.int/gaia}{http://www.cosmos.esa.int/gaia}), processed by the \textit{Gaia} Data Processing and Analysis Consortium (DPAC, \href{http://www.cosmos.esa.int/web/gaia/dpac/consortium}{http://www.cosmos.esa.int/web/gaia/dpac/consortium}). Funding for the DPAC has been provided by national institutions, in particular the institutions participating in the \textit{Gaia} Multilateral Agreement. This research has made use of the NASA Exoplanet Archive, which is operated by the California Institute of Technology, under contract with the National Aeronautics and Space Administration under the Exoplanet Exploration Program.

\software{
\package{matplotlib} \citep{hunter07},
\package{numpy} \citep{vanderwalt11},
\package{pandas} \citep{reback20},
\package{scipy} \citep{jones01},
\package{seaborn} \citep{waskom20},
}

\bibliographystyle{aasjournal}
\bibliography{mybib}

\end{document}